\documentclass[11pt]{article}

\usepackage{url} 
\usepackage[T1]{fontenc}
\usepackage{graphicx}
\usepackage{float} 
\usepackage{xcolor} 
\usepackage{lineno} 

\usepackage{amsmath,amssymb}
\usepackage{booktabs}
\usepackage{mathptmx}
\usepackage[margin=1in]{geometry}

\graphicspath{{figs/}{figures/}{pictures/}{images/}{./}}

\title{Evaluating Tactile Encoding Strategies for Communicating 3-D Shape}
\author{MacKenzie C. Harnett\\\small Texas A\&M University\\\small \texttt{mh2255@tamu.edu}
  \and Rebecca F. Friesen\\\small Texas A\&M University\\\small \texttt{rfriesen@tamu.edu}}
\date{}

\begin{document}
\maketitle

\begin{abstract}
Two-dimensional haptic displays, such as programmable pin arrays and vibrotactile or friction-modulating touchscreens, enable tactile communication of complex shapes in a low-profile form factor. However, many shapes of interest, including CAD models, three-dimensional graphs, and map topographies, are inherently three-dimensional. This work investigates two methods for encoding 3-D information using a 2-D tactile display: (1) a single tactile height map providing coarse but quantifiable depth cues, and (2) a series of binary layers communicating sequential slices of a 3-D model, supporting more complex geometries such as overhangs and cavities. We further compare performance across two display types—a friction-modulating display and a pin array—to examine how the level of haptic abstraction affects users’ ability to decode discrete 3-D information. A user study demonstrates that participants were able to effectively complete a 3-D assembly task using both strategies, achieving high accuracy across all display types. While the pin array supported faster task completion and higher-rated--though not statistically significant--user experience than the friction-modulating display for the binary-layer task, these differences did not translate to significant differences in overall usability (i.e., reassembly accuracy success). Together, these results identify design trade-offs and inform development of future surface haptic displays that support the communication and interpretation of discrete 3-D information in tactile graphics.

\end{abstract}


\vspace{1em}

\section{Introduction}

Touchscreens are an essential tool across professional, educational, and social contexts. 
Their 2-dimensional visual displays often communicate 3-dimensional data, including CAD models, topological maps, and multi-axis graphs, yet lack the rich haptic feedback provided by interacting with real-world objects and models.
Haptic technology offers an opportunity to both reintegrate meaningful tactile cues into digital tasks and expand the accessibility of touchscreen technologies.
Surface haptic displays, such as programmable pin arrays~\cite{humanware_website, dotpad_website, graphiti_website} and friction-modulating screens~\cite{tanvas_website}, provide tactile feedback on a surface form factor that could replace or augment a touchscreen's visual display. 

A core challenge for surface haptic displays is how to effectively convey 3-D tactile information through a 2-D tactile surface. 
Pin-array technology has been investigated for use in applications such as 3-D modeling and animation~\cite{Siu2019, Holloway2022}, while friction-modulating surfaces have been explored for multi-modal rendering of 3-D features~\cite{Kim2013, Osgouei2017}.
Much of this prior work, particularly in the latter case, uses haptics to \textit{enhance} visual feedback rather than convey 3-D structure independently. 
Each tactile rendering method has unique limitations; pin arrays provide physical height differences but are constrained by the low resolution of the physical pins, while friction-modulating displays can modulate friction at very high spatial resolution but lack the ability to render actual height differences that would directly communicate 3-D information. 
These limitations motivate the development of effective tactile encoding strategies that communicate 3-D information without necessarily reproducing physical geometry. However, limited work has systematically investigated the trade-offs between alternative tactile encoding strategies for communicating discrete 3-D structures using surface haptic displays. Addressing this gap could improve the information-encoding capabilities of surface haptic displays and expand the accessibility of digital graphics.

This study asks: how do different tactile encoding strategies influence the effectiveness of communicating 3-D shape information through surface haptic displays? 
To address this question, we evaluate two methods for conveying 3-D information through touch alone:
\begin{itemize}
\item \textbf{Tactile height maps}: Represent the height of each spatial region using one of several discrete tactile intensity levels on a single map. 
\item \textbf{Tactile binary layers}: Represent a 3-D structure as successive 2-D slices, each containing only two tactile states. 
\end{itemize}
A second, related question is: how do display characteristics, such as pin density and spatial segmentation, influence the effectiveness of these encoding strategies? To explore this question, we evaluated several variations of two surface haptic display technologies: pin arrays and friction-modulating displays. For the height map method, we compared two pin-array configurations with different pin densities and two friction-modulating configurations with and without a physical stencil to delineate spatial regions. For the binary layer method, we evaluated the standard pin-array and friction-modulating configurations. Physical 3-D models were also included as a reference condition.

To compare effectiveness of our different encoding strategies across display types, we assessed participants’ ability to interpret tactile stimuli on different displays and use the encoded information to reconstruct corresponding 3-D models. This required participants to distinguish between tactile stimulus categories and interpret their spatial arrangements to reproduce the corresponding structures.
Encoding effectiveness was evaluated through reconstruction accuracy, task completion time, hand movement, and subjective experience ratings. Together, these measures characterize both the accuracy of interpreting encoded information and the practical effort required to use each representation.

The successful integration of haptic tools into mainstream applications depends on both user experience and the practical utility of the information they provide. Rather than developing a new display, this work investigates how existing surface haptic technologies can be leveraged to communicate discrete 3-D shape information. By evaluating alternative encoding strategies and display configurations, we identify trade-offs in tactile information communication and opportunities to improve the design of future surface haptic displays. Ultimately, this work aims to inform the development of more effective tactile representations of 3-D information and expand the information-encoding potential of digital tactile graphics.

\section{Background}~\label{bckgrnd}
Our work is informed by broader research on haptic displays and technologies, their individual feedback mechanisms, and, more specifically, how these tools can be used in 2- and 3-D design frameworks. In the following section, we give particular focus on haptic materials as they are integrated into educational environments, as they represent a frequent and well-studied real-world application of such tools and technologies outside of niche research contexts. Here, we review the benefits and limitations of these technologies and their applications for rendering 3-D information. 

\subsection{Applications of Haptic Materials}~\label{sec:tactmatprac}
Haptic materials have shown promise in increasing engagement in remote education environments~\cite{richard2021}. This is accompanied by a growing interest in multimodal and modifiable audio-tactile materials for note-taking~\cite{Lee2023}, circuit diagrams~\cite{Davis2020}, and data visualization~\cite{Zhao2024, Ramoa2023, Engert2022, Borah2022, Fan2022}, all of which can particularly benefit blind and low vision (BLV) students. Note-taking and data visualization are essential for participating in mainstream education curricula; however, in practice, accessible tactile materials are often limited by high tooling requirements, storage costs, and immutability~\cite{Phutane2022}. These constraints negatively affect the incorporation of multimodal materials into classroom environments. 
To address these limitations, it is necessary to develop haptic tools that reduce storage and cost burdens, while also creating pipelines and systems through which both sighted and BLV students can effectively use, interpret, and engage with these materials~\cite{Borah2022}.
Despite their potential, current haptic materials remain underutilized in educational settings due to these practical barriers.

\subsection{Refreshable Haptic Displays}

\subsubsection{Virtual, Augmented, and Mixed Reality (XR) Systems} 
There is significant interest in integrating haptic feedback into XR systems to enhance embodiment and engagement~\cite{richard2021}. Several of the haptic actuation methods described below have been incorporated into XR systems at the fingertip level--including vibrotactile motors~\cite{richard2021, Normand2025, Elsayed2020}, electro-tactile stimulation~\cite{Suga2023}, and shape displays~\cite{Kourtesis2022}. Force feedback is also a common integration, especially in XR wearables~\cite{Suga2023, Yano2015, Wang2023}. Applications for haptic technologies in virtual reality environments include 2-D sketching~\cite{Mohanty2019, Elsayed2020}, dynamic room configurations for immersive environment design~\cite{Suzuki2020, Je2021}, and fostering social touch~\cite{Talhan2024, Hecquard2023, Maunsbach2023}. Notably, haptic interactions in virtual reality environments do not need to be fully realistic--they only need to be intuitive~\cite{Rosenkranz2024}, which indicates promise for more 'abstract' representations of surfaces when using haptic feedback. 
While XR platforms may better render 3-D surfaces intuitively through haptic feedback, surface haptic displays offer unique advantages: they are more accessible, portable, and provide a stable surface for touch-based interactions. These qualities make them particularly suitable for widespread adoption.

\subsubsection{Shape Displays}  
Shape displays come in many forms, including bubble arrays driven by pneumatic~\cite{Shan2024, Talhan2024} or electroosmotic~\cite{Grant2024, Shultz2023, Shen2023} pumps, electromagnetic pin arrays~\cite{Chen2023, Yasu2024}, and mechanically-driven pin arrays~\cite{graphiti_website, humanware_website}. Generally, smaller, low-resolution arrays are incorporated into XR wearables, while larger platforms are used for more diverse design-based applications, some of which are described in the following sections. 

This work focuses on mechanically driven pin arrays, one of the most developed shape display technologies. Commercially available devices~\cite{graphiti_website, humanware_website, dotpad_website} consist of arrays of pins with programmable heights, which can be actuated to create graphics or objects with discernible edges or geometries. These displays have been used in a variety of applications, including animation~\cite{Holloway2022}, sketching~\cite{Bornschein2018}, and 3-D modeling~\cite{Siu2019}. Compared to alternative haptic displays, these are the most well-established for the nature of work described in this paper. Unfortunately, they tend to be prohibitively expensive--a barrier to wide dissemination~\cite{Mukhiddinov2021}; however, recent research has focused on developing affordable alternatives~\cite{Chen2023, Yasu2024}. 

\subsubsection{Friction-Modulating and Vibrotactile Displays} 
Friction modulation functions by varying the friction forces between the fingerpad and a surface, primarily through electroadhesion~\cite{bau2010, vardar2017, shultz2018, meyer2014} or ultrasonic friction reduction~\cite{wiertlewski2016, biet2007, watanabe1995}. These surfaces can render dynamic textures and interactions~\cite{friesen2021}. In the past, these tools have been commercially available~\cite{tanvas_website} at a relatively affordable price point (\$1999 USD) compared to other refreshable surface haptic displays. 

One of the most common haptic integrations in mainstream technology is the use of vibrotactile motors. Similar to friction-modulating screens, vibrotactile motors are associated with 'whole-screen' feedback. Recent work has shown the potential for high-resolution, dynamic, and localized feedback using vibrotactile motors; however, these configurations require a high-density of peripheral electronics~\cite{reardon2023} which limits scalability. Vibrotactile displays have also been used for texture rendering~\cite{Friesen2024, Normand2025, Strohmeier2017} and dynamic surface interactions such as button clicks~\cite{Sadia2020}. 
Both friction modulation and vibrotactile feedback provide avenues for representing surface information in ways that differ from discrete pin heights, making them relevant for exploring alternative 3-D rendering strategies in surface haptic displays.

\subsection{Utilizing Haptic Tools for Accessible Design}~\label{sec:tactmatdesign}

2- and 3-D sketching are essential communication tools~\cite{Goel1995, Goldschmidt1994, Suwa1997} and support cognitive development~\cite{Goel1995} and ideation~\cite{Tversky2003}. 
Engaging students--including BLV students--in these practices can be facilitated by integrating multimodal feedback into classroom materials. The following sections review the applications and limitations of haptic tools in 2- and 3-D design.

\subsubsection{2-D Design}
A popular integration for haptic displays is to facilitate accessible and multimodal 2-D design pipelines. Traditional methods for creating tactile sketches are often limited due to the immutability of the resultant images. Refreshable displays offer an opportunity to create infinitely reproducible, distributable, and highly storable tactile graphics. 
Existing haptic tools for creating refreshable tactile sketches include force~\cite{Mohanty2019} or vibrotactile~\cite{Elsayed2020} feedback in XR environments; modular blocks that are later scanned and stored as 2-D graphics~\cite{Bornschein2018}; and shape displays that support pen- and hand-based drawings~\cite{Bornschein2018, Harnett2025}. Alternative tactile tools remain promising, but unrealized, in generating accessible 2-D graphics. 
For example, friction-modulating displays struggle to render shape primitives and basic geometric features~\cite{Gershon02016, Osgouei2016, Rohou2025, Sadia2022, Harnett2025_2}, and pin arrays are limited by resolution, making it difficult to distinguish higher-order polygons~\cite{Bhatnagar2023,O’Modhrain2015}. These limitations reduce both the complexity and saliency of designs rendered using tactile feedback. Emerging approaches, such as AI-assisted workflows~\cite{Bennet2024} or grammar-based design tools~\cite{Ohene-Djan2016}, do not inherently provide tactile feedback but could be leveraged within more mature tactile design systems.

\subsubsection{3-D Design}
Visual methods for conveying 3-D information as 2-D projections (e.g., perspective or isometric drawings) are inaccessible to BLV individuals because they generally cannot be coherently rendered using pins or raised line drawings~\cite{Panotopoulou2020}. Producing physical 3-D tactile materials can be expensive, inefficient, and time-consuming, as noted in Section~\ref{sec:tactmatprac}. This creates an open challenge: how can 3-D CAD pipelines be made more engaging, accessible, and efficient for a broader community of designers?

Some groups that have integrated interactive shape displays into the 3-D design process~\cite{Siu2019, Schoessler2015} have developed methods of digitizing user-made, physical structures created with modular components~\cite{Lee2024}; and others have used AI to control shape-changing displays using 'text-to-shape prompts'~\cite{Qian2024}.
Other refreshable displays, such as friction-modulating surfaces, have been used to enhance 3-D visuals rather than conveying 3-D information independently~\cite{Kim2013}, which they struggle to do without guidance or context~\cite{Osgouei2017}. 

Graphic design strategies can support 3-D communication via touch. When interacting with 2-D geometries, tactile readers often interpret structures sequentially, building a whole from parts~\cite{Jiang2025, Bhatnagar2023}. This sequential interpretation of 2-D information is also seen in the interpretation of 3-D information. A similar approach is effective for 3-D structures; for example, BLV individuals report positive experiences when interpreting internal building structures as a series of overlapping "slices"~\cite{Nagassa2023}. While visual perspective drawings cannot be rendered 1:1 on tactile materials, simple structures are generally coherent~\cite{Shimojo1997}, and more complex structures can be represented as part-based projections that emphasize key features over strict realism~\cite{Panotopoulou2020}. 

While previous work has provided promising tools for virtual haptic 3-D shape rendering~\cite{Hirota2026, Subramanian2014, Shinoda2020}, especially in a VR context, relatively little work has investigated the utility of or compared different refreshable, portable surface haptic displays in their ability to convey 3-D surface information. Previous work on multi-height diagrams and textures has focused on non-digital materials~\cite{Gupta2019} and did not address the mechanical limitations or unique properties of refreshable haptic displays.

\subsection{Our Work}
As discussed in Section~\ref{bckgrnd}, the information-encoding capabilities of surface haptic displays remain largely underutilized, particularly for conveying 3-D information. This work focuses on \textit{surface} haptic displays due to the ubiquity of touchscreens and the relative ease of generating content for them compared with XR-based tactile graphics.
Although a range of commercially available surface haptic displays exists, it is unclear how effectively their feedback modalities can be leveraged to communicate depth and how their designs might be refined to address current limitations. For example, increasing the pin density of a pin array may support more continuous surface features and reduce interruptions in tactile exploration. However, its impact on users’ ability to interpret and reconstruct 3-D information has not been systematically evaluated.

We hypothesize that the two proposed rendering methods will be better suited to different display types. For the height map method, more abstract encodings—such as texture keys on friction-modulating displays—may provide a promising approach for encoding quantitative 3-D information. Friction-modulating displays can produce a near-unlimited set of distinct textures, enabling richer representations than the discrete height levels available on pin arrays. In contrast, pin arrays produce crisp edges that delineate object boundaries, but their mechanical actuation limits the achievable height range and z-resolution of rendered surfaces. Texture patterns on friction-modulating displays can also be interpreted independently, whereas pin heights are typically judged relative to a reference point.

Conversely, we expect that the binary layer method will leverage haptic cues more efficiently, making it particularly advantageous for pin arrays with limited or binary actuation heights. Although friction modulation can produce a range of discernible textures, generating reliably distinct variations along a single perceptual dimension is challenging. As a result, the texture-based approach may be less effective than raised pin geometries for rendering 3-D information with the binary layer method.

To evaluate these hypotheses, we designed a user study comparing the proposed 3-D encoding methods across immutable reference models (e.g., 3-D prints), commercially available displays (e.g., Tanvas), commercial proxies (e.g., Graphiti), and modified display configurations that highlight each device’s strengths and explore opportunities for improving information encoding and communication.

\section{3-D Encoding with Tactile Height Maps}

\subsection{Experimental Configuration}

\begin{figure}[htbp!]
\centering
\includegraphics{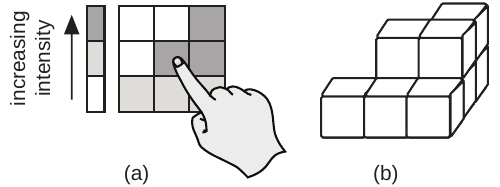}
\caption{\textbf{Height Map Rendering Method Schema}. (a) The stimuli intensity map, where shade corresponds to the intensity of output \textit{and} a discrete height value, and (b) the associated final assembly.}
\label{fig_hmschema}
\end{figure}

Our first rendering method, the tactile \textit{height map}, presents a model as an n-by-n matrix where each cell has a tactile key corresponding to a discrete height value (0, 1, or 2) of a reference model (Figure~\ref{fig_hmschema}).

The experimental configuration consisted of one of five display configurations: a 3-D-printed reference model, a low-density pin array, a high-density pin array, a friction-modulating display without a stencil, or a friction-modulating display with a stencil. These configurations represented two refreshable surface haptic display architectures (pin arrays and friction-modulating displays), alongside a physical reference condition.

\begin{figure}[htbp!]
\centering
\includegraphics{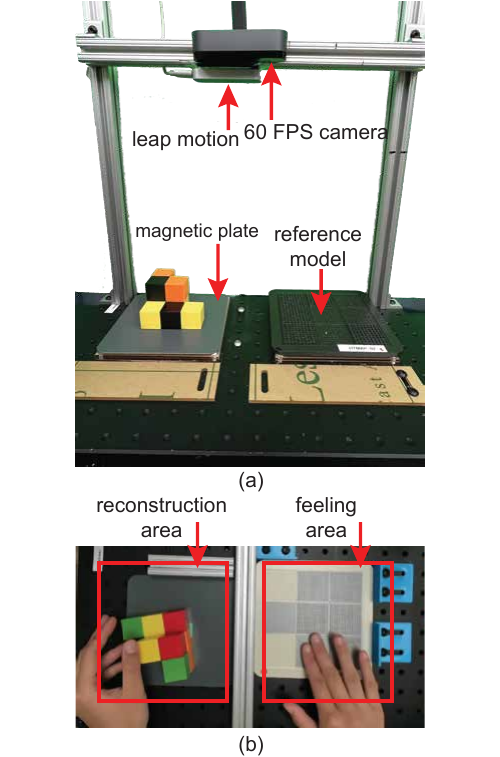}
\caption{\textbf{Experimental Setup for Height Map Assembly Task}. The (a) front view and (b) top-down view of the final experimental configuration for the tactile "height map" assembly task. The experimental setup is obscured behind an opaque curtain during data collection.}
\label{fig_expsetup}
\end{figure}

Beyond the actual interactive displays, it included a hand tracking system and an assembly zone consisting of a magnetic plate and twenty-seven magnetic blocks. All tasks were obscured behind an opaque curtain to ensure participants relied solely on tactile feedback. Figure~\ref{fig_expsetup} illustrates the final setup.


\begin{figure}[t!]
\centering
\includegraphics{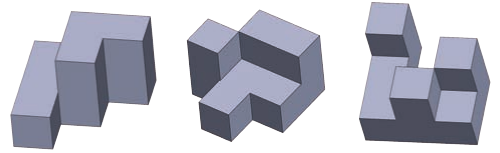}
\caption{\textbf{Sample Models for Height Map Assembly Task}. These three different models constituted the three trials of each assembly task. These same models were evaluated for both the friction modulation and pin array sub-experiments.}
\label{fig_samplesHM}
\end{figure}

\subsubsection{Reference Models} 
~\label{section:hapticdisplaytypes}

\begin{figure}[t!]
\centering
\includegraphics{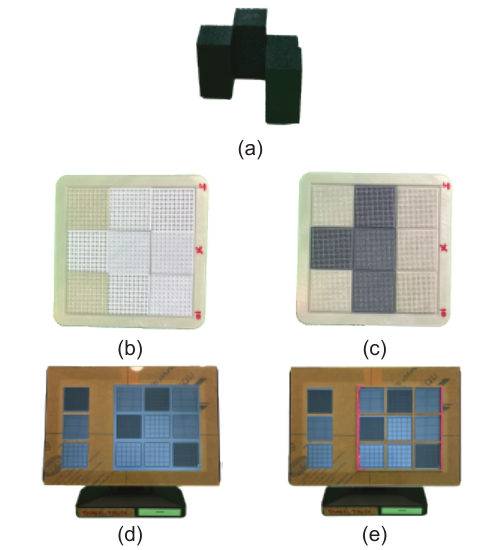}
\caption{\textbf{Display Types for Height Map Assembly Task}. These three display types were evaluated for the 'tactile height map' method. (a) 3-D printed model, (b) low-density pin array, (c) high-density pin array, (d) friction-modulating display without a stencil, (e) friction-modulating display with a stencil.}
\label{fig_methodsHM}
\end{figure}

Three 3-D models were created, each consisting of nine $1~in^3$ sections with heights of 0, 1, or 2 cubes. Heights were equally and randomly distributed across a 3-by-3 matrix. These matrices served as templates for reference models made of different display types, including pin arrays, friction-modulated texture grids, and 3-D prints. We then observed participants' ability to reconstruct the reference models when presented on the different display types. Limiting models to simple 3-by-3 structures simplified analysis, ensured comparable difficulty, and minimized participant fatigue (Figure~\ref{fig_samplesHM}).
For each display type, tactile intensity was mapped to the three height values. For pin arrays, this corresponded to a height range of 0.5–1.5 mm, while for friction-modulating displays, intensity mapped to spatial frequencies of on-screen textures. All five display types are visualized in Figure~\ref{fig_methodsHM}.

The two types of pin array displays were first evaluated against each other. These included: 

\begin{itemize}
\item \textbf{Low-Density Pin Array:} 3-D printed static display with pins spaced 4.1 mm apart, heights 0.5–1.5 mm.

\item \textbf {High-Density Pin Array:} Same as low-density, but pin density doubled.
\end{itemize}

The low-density configuration was designed to represent the dimensions and pin spacing of the commercial Graphiti display. A static implementation was used rather than the refreshable hardware due to reliability issues and a tendency for the pins to malfunction during operation. This allowed us to maintain consistent stimulus configurations across participants and trials. The higher-density configuration was designed to explore whether reducing the spacing between pins could provide a more continuous tactile surface and facilitate exploration of the height map.

\noindent
Two types of friction-modulating display were compared to each other in a separate session. Since friction-modulated textures are more disambiguated from height information, this session also included a control comparison with a 3-D physical model. 

\begin{itemize}
\item \textbf{3-D Printed Model:} Control to compare user performance and enjoyment. Durable but difficult to store or manufacture. 
\item \textbf{Friction-Modulating Display:} Commercially available display with textures mapped to block heights. Roughness (i.e., spatial frequency) is the adjusted intensity value. 
\item \textbf{Sectioned Friction-Modulating Display:} Same as above friction-modulating display, but with a stencil dividing each 3-by-3 cell into discrete sections. 
\end{itemize}

Textures for the friction-modulating display were designed using the Weber fraction for discriminating roughness. We used a Weber fraction value of 0.19~\cite{Tymms2018}. The maximum spatial frequency was 0.60 $px^{-1}$, with intermediary and minimum frequencies calculated as:

\begin{equation}
    x_i= x_{i-1}*1.19^7
\end{equation}

The Weber value was raised to the power of 7 to ensure that the difference between the textures was perceivable while still scalable to the dimensions of the friction-modulating display. Values were manually adjusted to ensure they could be adequately distributed within the allocated space. This calculation produced a texture set corresponding to spatial frequencies of .60, .1745, and .05 $px^{-1}$. 


\subsubsection{Hand Tracking}~\label{handtrack_description}
We used a 60 FPS camera (Logitech 960-001178) mounted directly above the work area to record time-to-completion and identify unique exploration behaviors. We also implemented dual-hand tracking at 120 FPS with a Leapmotion Controller (Ultraleap) to estimate relative energy expenditure during assembly tasks.

\subsection{Participant Study}

Participants were asked to explore tactile stimuli presented on a static display and then reconstruct the corresponding 3-D model.
This portion of the study was divided into two parts, each comparing different approaches to using a particular display type: (1) a comparison between two pin array configurations, and (2) a comparison between two friction-modulating displays and a reference 3-D model (see Section~\ref{section:hapticdisplaytypes}).
Fifteen participants (6 women, ages 19–34, all right-handed) took part in the pin array comparison, and a separate group of fifteen participants (6 women, ages 19–50, all right-handed) participated in the friction-modulation and 3-D model comparison. Each participant completed only one part of the study, and each study session lasted approximately one hour.
This study was approved by the Institutional Review Board of Texas A\&M University (IRB2024-0539). All participants provided informed consent prior to participation and were compensated for their time.


\subsubsection{Familiarization Session}~\label{sec:training_HM}
Before data collection, participants first completed a familiarization session to assess their tactile acuity and increase their familiarity with the displays. 
For acuity assessment, participants sorted three variable-spatial-frequency textures in order of intensity on a friction-modulating display across three randomized repetitions. Sorting at least two-thirds correctly was considered passing; however, this assessment was not used to determine participant eligibility, and all participants proceeded regardless of performance.
Next, participants were asked to interpret a reference model on each display type and assemble it as they would in the main experiment. 
During this familiarization, they were instructed to keep their hands under the curtain and use the index finger of their dominant hand when touching the friction-modulating display. For 3-D printed models, they were instructed not to detach the model from its magnetic mount. Pink noise via headphones prevented auditory cues from influencing texture identification. Participants were able to freely move between the display and the blocks throughout the familiarization task, allowing them to refer back to the tactile stimuli while assembling the reference model. They were not told at any point whether they had successfully built a model. 

\subsubsection{Assembly Task}
Participants interpreted and recreated three 3-D models for each display in pseudo-random order. Throughout all tasks, their hands remained under the curtain for uninterrupted tracking. They reassembled a set of magnetic blocks with unit height values corresponding to the intensity of stimuli rendered on the static display (i.e., pin height for the pin array, spatial frequency for the friction-modulating display). Unlimited time was allowed for each assembly, followed by a one-minute break. Participants were not provided with correctness feedback during the assembly task. Accuracy was calculated after each assembly was completed.


Participants were able to attain a maximum assembly 'score' of 15. Confusing the highest and lowest frequency textures was considered a more severe error than confusing either extreme with an intermediary texture. In this study, the former error type led to a deduction of 2 points, and the latter a deduction of 1 point. The participant's final score was then divided by the maximum possible score, and the resultant calculation constituted each trial's assembly accuracy: 

\begin{equation}
Accuracy~(\%) = N/15.0
\end{equation}~\label{eqn:acc1}

\noindent
After each block, participants completed a survey and took a three-minute break before the next task.

\subsubsection{Interaction Method Survey}
After each block, participants were asked to complete a survey rating their experience. A 10-point Likert scale was used, where a 1 corresponded to 'Not at all Difficult' or 'Not at all Confident' and a 10 corresponded to 'Very Difficult' or 'Very Confident'. A 10-point scale was chosen so that the data could later be treated as continuous. This questionnaire is presented in Table~\ref{table:questions}. 

\begin{table}[!t]
\renewcommand{\arraystretch}{1.35}
\caption{Participant Survey}
\label{table:questions}
\centering
\begin{tabular}{|c|p{6 cm}|}
\hline
\textbf{No.} & \textbf{Statement} \\
\hline
Q1 & Rate, on a scale of 1-10, the difficulty of the completed assembly task. \\
\hline
Q2 & Rate, on a scale of 1-10,  your confidence in recreating the rendered model. \\
\hline
Q3 & Rate, on a scale of 1-10,  your enjoyment of the assembly task. \\
\hline
\end{tabular}
\end{table}

\section{3-D Encoding with Tactile Binary Layers}

\subsection{Experimental Configuration}
\begin{figure}[t!]
\centering
\includegraphics{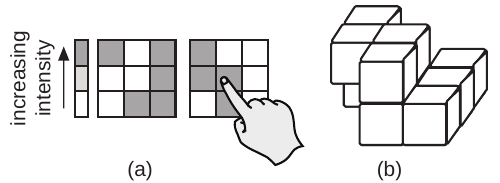}
\caption{\textbf{Binary Layer Rendering Method Schema}. (a) The stimuli intensity maps, where shade corresponds to the intensity of output \textit{and} a discrete height value, and (b) the associated final assembly.}
\label{fig_blschema}
\end{figure}

The binary layer method evaluated in this second experiment encodes 3-D information as a \textit{sequence of binary intensity layers}. Each layer is an n-by-n matrix where each cell corresponds to one of two discrete height values in the reference model. This reduces the number of unique intensity values while enabling representation of complex features such as overhangs or internal structures (Figure~\ref{fig_blschema}).

\begin{figure}[t!]
\centering
\includegraphics{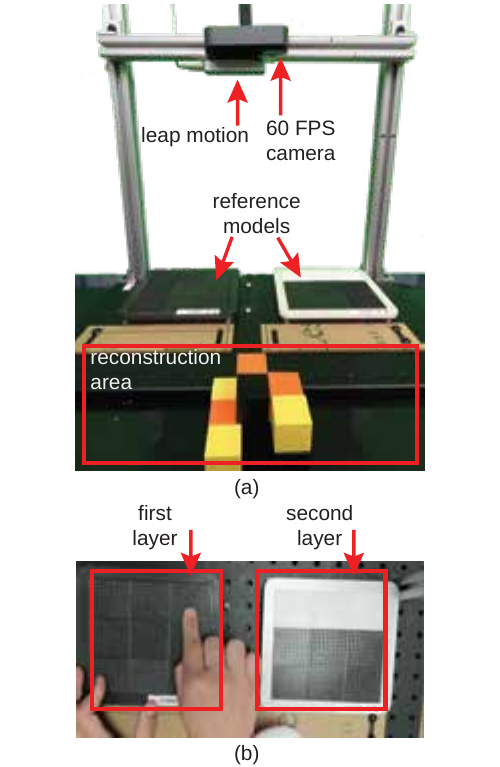}
\caption{\textbf{Experimental Setup for Binary Layer Assembly Task}. The (a) front view and (b) top-down view of the final experimental configuration for the tactile "binary layer" assembly task. The experimental setup is obscured behind an opaque curtain during data collection.}
\label{fig_expsetup_bl}
\end{figure}

The experimental configuration comprised one of three display conditions: a 3-D printed model, a pin array, or a friction-modulating display; a hand tracking system; and eighteen magnetic blocks. The refreshable surface haptic displays (pin arrays or friction-modulating) were presented on two separate surfaces to simulate a sequence of information, allowing participants to move back and forth between layers. Tasks were performed behind an opaque curtain to ensure participants relied solely on tactile feedback. Figure~\ref{fig_expsetup_bl} illustrates the configuration. Here, we only tested the two configurations that more closely reflect the typical operating conditions of the associated surface haptic displays. This was because our primary focus was evaluating how 3-D information could be rendered as a sequence of information across different display types, rather than optimizing individual display configurations.

\begin{figure}[t!]
\centering
\includegraphics{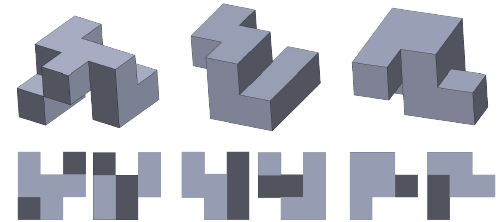}
\caption{\textbf{Sample Models for Binary Layer Assembly Task}. These three different models constituted the three trials of each assembly task. Projections of the tops and bottoms of the surface are included to emphasize the presence of overhangs in the models.}
\label{fig_samplesBL}
\end{figure}

\subsubsection{Surface Haptic Display Types}

\begin{figure}[t!]
\centering
\includegraphics{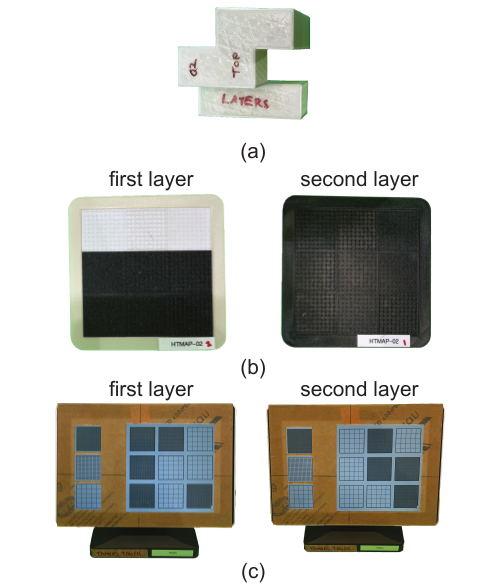}
\caption{\textbf{Display Types for Binary Layers Assembly Task}. These three displays were evaluated for the "binary layers" method. These consist of a (a) 3-D printed model, (b) two pin arrays, and (c) two friction-modulating displays.}
\label{fig_methodsBL}
\end{figure}

Three 3-D models were created, each consisting of nine $1~in^3$ sections with heights of 1 or 2 cubes. The “on”/“off” states were randomly distributed across two 3-by-3 matrices, which were layered to generate each reference model for CAD and 3-D printing (Figure~\ref{fig_samplesBL}).

For each display, tactile intensity was mapped to the two height values. In pin arrays, heights were 0.5 or 1.5 mm; in friction-modulating displays, spatial frequencies were 0.05 or 0.6 $px^{-1}$. Each display type is described below: 

\begin{itemize}
\item \textbf{3-D Printed Model:} A control for evaluating performance and enjoyment of the other displays. Durable but difficult to store and manufacture. 
\item \textbf{Friction-Modulating Display:} Two commercially available friction-modulating displays. Textures with variable spatial frequencies map to each block height value.
\item \textbf{Pin Array:} A set of two static 3-D printed displays based on the dimensions of existing commercial pin displays. Pin heights are 0.5 or 1.5 mm, and each pin is spaced approximately 2.05 mm apart center-to-center.
\end{itemize}

\noindent
These display types are shown in Figure~\ref{fig_methodsBL}. 

\subsubsection{Hand Tracking}
Identical to the tracking described in the height map experiments, which are described in Section~\ref{handtrack_description}.

\subsection{Participant Study}
A total of 15 people (6 women, ages 20-34, all right-handed) participated. Each study session was approximately 45 minutes to 1 hour in length.
This study was approved by the Institutional Review Board of Texas A\&M University (IRB2024-0539), and all participants signed an informed consent form before data collection and were compensated for their participation.

\subsubsection{Familiarization Session}~\label{sec:training_BL}

This familiarization session mirrored Section~\ref{sec:training_HM}, with adjustments for binary layers. Participants completed a tactile acuity assessment with three textures, randomly scrambled, and rated each as Low, Medium, or High. Passing required at least two-thirds correct, but all participants continued regardless of results.
They then reassembled a reference model for each display, during which they kept hands under the curtain, used only the dominant-hand index finger for friction-modulating displays, and did not detach 3-D models from magnetic mounts. Pink noise blocked auditory cues.
Participants explored one display at a time and could freely move between the displays and their assembly throughout the familiarization task, allowing them to refer back to the tactile information corresponding to each layer. Once familiar with all displays, the main task began.

\subsubsection{Assembly Task}
Participants were asked to 
recreate a set of three 3-D models for each display configuration within an experimental set. The displays and models were presented in a pseudo-random order. Participants were reminded to keep their hands under the curtain. 
When recreating block models from haptic display information, magnetic blocks corresponded to tactile stimuli (pin height or spatial frequency), while for the 3-D model, participants reassembled a single complete model. Unlimited time was allowed for each assembly, followed by a one-minute break. Participants were not provided with correctness feedback during the assembly task. Accuracy was calculated after each assembly was completed.

Participants were able to attain a maximum assembly 'score' of 9 for each layer (18 total). A misidentified frequency resulted in a deduction of 1 point. The participant's final score was then divided by the maximum possible score, and the resultant calculation constituted each trial's assembly accuracy: 

\begin{equation}
Accuracy~(\%) = N/18.0
\end{equation}~\label{eqn:cc2}

At the end of each experimental block, participants were asked to complete a survey and given a three-minute break before proceeding to the next assembly task. 

\subsubsection{Interaction Method Survey}
At the end of each experimental set, participants were asked to complete a survey to rate their experience when assembling a 3-D model. This questionnaire, identical to the height map questionnaire, is presented in Table~\ref{table:questions}. 

\section{Results}

\subsection{Rendering with Height Maps: Performance}
When applicable, we assessed normality using a Shapiro–Wilk test, followed by two-way repeated measures ANOVAs examining the effects of display and model type on performance. Post-hoc paired-sample t-tests with Bonferroni corrections were used for pairwise comparisons.
\begin{figure}[H]
\centering
\includegraphics{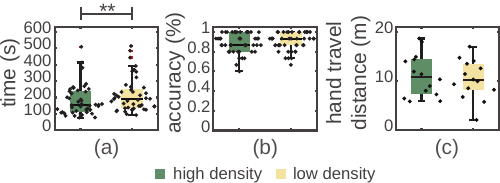}
\caption{\textbf{Performance Results for Pin Arrays in Height Map Assembly Task.} (a) The total time for a participant to complete each assembly task for both of the pin array display configurations; (b) accuracy was measured by calculating the number of correctly placed blocks divided by the maximum possible score; and (c) total distance traveled by each of the hands for each display type. Data shown is only for participants where hand tracking was measured as 'successful'. This data is presented for both of the pin array display configurations. ** corresponds to p $<$ 0.01.}
\label{fig_HM  PA All}
\end{figure}


\subsubsection{Accuracy on Pin Arrays}

Accuracy was averaged across the three trials. No significant differences were observed between low- and high-density pin arrays (p = 0.487 > 0.05). Data transformation using a Tukey Ladder of Powers (lambda = 2.625) was applied prior to analysis. Untransformed data is shown in Figure~\ref{fig_HM  PA All}(b).


\subsubsection{Time-to-Completion on Pin Arrays} 

Time-to-completion was defined as the total time, in seconds, required to assemble and submit the model. Participants completed assemblies significantly faster on the high-density pin array than on the low-density configuration (p = 0.006). A Tukey Ladder of Powers produced a lambda value of -0.45, and each value in the original data set was raised to that power and multiplied by -1 before analysis; untransformed results are shown in Figure~\ref{fig_HM  PA All}(a).


\subsubsection{Energy Expenditure on Pin Arrays}

Energy expenditure was calculated as the total distance traveled by participants' hands during each assembly task. This was averaged across all three trials for each condition. A one-way ANOVA was used to assess the impact of display configuration, which found that platform (high or low density pins) did not have a significant effect on energy expenditure (p=0.435$>$0.05).
A Tukey Ladder of Powers was conducted on the performance data, and a data transformation (lambda = 0.925) was applied prior to analysis; see Figure~\ref{fig_HM  PA All}(c). 
Because the quality of hand tracking data varied between participants, trials where hand tracking 'failed' (i.e., $<$ 20\% tracking success) were excluded from analysis. 

\begin{figure}[H]
\centering
\includegraphics{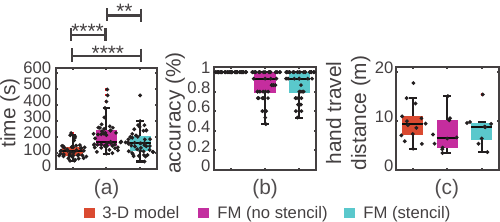}
\caption{\textbf{Performance Results for 3-D Model and Friction-Modulating Screens in Height Map Assembly Task.} (a) The total time for a participant to complete each assembly task for all three configurations; (b) accuracy was measured by calculating the number of correctly placed blocks divided by the maximum possible score; and (c) total distance traveled by each of the hands for each display type. Data shown is only for participants where hand tracking was measured as 'successful'. ** corresponds to p $<$ 0.01, **** corresponds to p $<$ 0.0001.}
\label{fig_HM FM All}
\end{figure}


\subsubsection{Accuracy on Friction-Modulation, 3-D Models}

Accuracy analysis was complicated by participants achieving near-perfect performance on the 3-D model; therefore, no ANOVA was performed.
There was also no significant difference in performance between participants who failed the acuity task and those who passed; thus, the pool of participants is presented in its totality. Data are visualized in Figure~\ref{fig_HM FM All}(b). 


\subsubsection{Time-to-Completion on Friction-Modulation, 3-D Models}


There was a significant difference in time between the 3-D model and unsegmented friction-modulating display (p=3.36E-11$<$0.05), the 3-D model and the segmented screen (p=4.08E-05$<$0.05), and between the two friction-modulating screens (p=1.00E-02$<$0.05) for assembling the rendered models. Participants were able to complete the assembly task fastest when using the 3-D model, followed by the segmented screen. Completion times were at their highest when using the friction-modulating screen without a stencil. 
A Tukey Ladder of Powers was conducted on the performance data; data were transformed with lambda = -0.1 (multiplied by -1) prior to analysis. Figure~\ref{fig_HM FM All}(a) shows untransformed results.

\subsubsection{Energy Expenditure on Friction-Modulation, 3-D Models}
Energy expenditure was measured through the total distance traveled by participants' hands during each assembly task. This was averaged across all three models. A one-way ANOVA was used to assess the impact of display configuration, which found that platform did not have a significant effect on energy expenditure (p=0.467$>$0.05).
A Tukey Ladder of Powers was conducted on the performance data. This produced a lambda value of 0.15, and each value in the original data set was raised to that power before analysis. The untransformed data is visualized in Figure~\ref{fig_HM FM All}(c).
Because hand tracking varied between participants, trials where hand tracking was assessed to have 'failed' (i.e., $<$ 20\% tracking success) were excluded from analysis. 

\subsection{Rendering with Height Maps: User Experience}

\begin{figure}[H]
\centering
\includegraphics{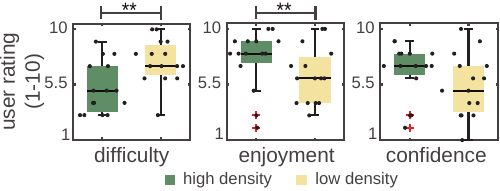}
\caption{\textbf{User Experience Ratings for Pin Arrays in Height Map Assembly Task.} Participants were asked to rate, on a scale of 1-10: (A) how difficult the assembly task was, (B) how much they enjoyed the assembly task, (C) how confident they were in interpreting and assembling the 3-D Model. ** corresponds to p $<$ 0.01.}
\label{fig_PA Survey}
\end{figure}

\subsubsection{User Experience for Pin Arrays}
A one-way ANOVA was used to assess the impact of display type on user experience, and a post-hoc paired sample t-test with a Bonferroni correction was used to assess the relative effect of each. There was a significant difference in perceived task difficulty (p=0.003$<$0.05), where the high-density array task was perceived as easier; in task enjoyment (p=0.007$<$0.05), where the high-density array was perceived as more enjoyable; but not in confidence (p=0.054$>$0.05) when building the final assembly. 
Tukey Ladder of Powers transformations were lambda = 1.4, 1.5, and 1.5 for difficulty, enjoyment, confidence.
The untransformed data is visualized in Figure~\ref{fig_PA Survey}.

\begin{figure}[H]
\centering
\includegraphics{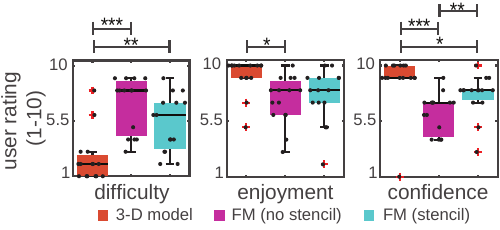}
\caption{\textbf{User Experience Ratings for 3-D Model and Friction-Modulating Screens in Height Map Assembly Task.} Participants were asked to rate, on a scale of 1-10: (A) how difficult the assembly task was, (B) how much they enjoyed the assembly task, (C) how confident they were in interpreting and assembling the 3-D Model. * corresponds to p $<$ 0.05, ** corresponds to p $<$ 0.01, *** to p $<$  0.001.}
\label{fig_FM Survey}
\end{figure}

\subsubsection{User Experience for Friction-Modulation, 3-D Models}
There was a significant difference in perceived task difficulty between the 3-D model and unsegmented friction-modulating display (p=1.88E-4$<$0.05), with the 3-D model outperforming the unsegmented surface; the 3-D model and segmented surface (p=8.00E-03$<$0.05), with the 3-D model outperforming the segmented surface; but not between the unsegmented and segmented surface (p=0.05=0.05). There was a significant difference in task enjoyment between the 3-D model and unsegmented friction-modulating display (p=0.012$<$0.05), with the 3-D model outperforming the unsegmented display; but not between the 3-D model and segmented surface (p=0.054$>$0.05) or the unsegmented and segmented surface (p=1$>$0.05). There was a significant difference in confidence between the 3-D model and unsegmented friction-modulating display (p=0.0002$<$0.05), the 3-D model and segmented surface (p=0.027$<$0.05), and between both friction-modulating screens (p=0.006$<$0.05), Confidence was at its highest when using the 3-D model, followed by the segmented surface. It was at its worst when using the unsegmented friction-modulating display. 

Tukey Ladder of Powers transformations were lambda = 2.35, 1.5, and 2.375 for difficulty, enjoyment, and confidence. The untransformed data is visualized in Figure~\ref{fig_FM Survey}.


\subsection{Rendering with Height Maps: Exploration Strategies}

\begin{figure}[H]
\centering
\includegraphics{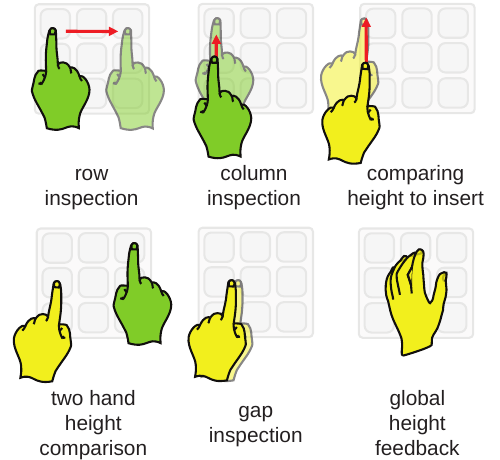}
\caption{\textbf{User Exploratory Behaviors on Static Pin Arrays}. For each display type, the darker shaded hand corresponds to the dominant hand, and the lighter shade corresponds to the non-dominant hand. Semi-transparent hands correspond to the final position of a hand in motion. Arrows correspond to the path of a traveling hand.}
\label{fig_exppin}
\end{figure}

Non-quantifiable behaviors were observed between participants when exploring the surface displays. These variable behaviors may or may not have significantly impacted task performance. Selected participant comments were recorded to help interpret exploratory behaviors or experiences. 

\subsubsection{Exploration of Pin Arrays}
One participant noted that a gap made the task more difficult because their finger would get stuck between them during exploration and disrupt their relative height assessments, specifically for the high-density pin array. This participant rated enjoyment lower and difficulty higher and performed worse on the high-density array than the low-density array, which was in direct contrast with other participants. Another participant compared the high-density array to having a 'blindfold removed', and they did comparatively better on the high-density array versus the low-density array. 

Another participant noted that the amount of information they could collect traveling vertically differed from when traveling horizontally; however, they did not note \textit{which} strategy produced more information. This may indicate that individual exploration strategies impact performance and perception. Participants expressed that the low-density pin array was more abrasive to the fingerpad than the high-density array--this makes sense because forces are more evenly distributed across the fingerpad for the high-density array. However, these experiences did not affect confidence or performance during assembly tasks. Selected exploration strategies are recorded in Figure~\ref{fig_exppin}. 

\subsubsection{Exploration of Friction-Modulation}

\begin{figure}[H]
\centering
\includegraphics{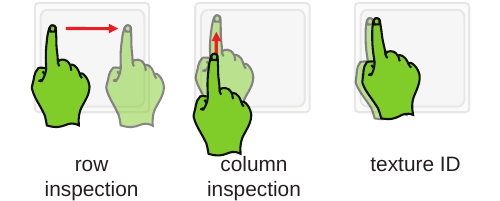}
\caption{\textbf{User Exploratory Behaviors on Static Friction-Modulating Display}. For each display type, the shaded hand corresponds to the dominant hand. 
Semi-transparent hands correspond to the final position of a hand in motion. Arrows correspond to the path of a traveling hand.}
\label{fig_exppin2}
\end{figure}

Participants were limited in the types of exploratory behaviors they could use on a friction-modulating display because exploration was restricted to a single finger. This prevented simultaneous comparisons between sections of the grid. When using the friction-modulating display with no stencil, a participant commented that, by identifying individual textures instead of scanning the entire expanse of the screen, they felt better able to perform the assembly task, which was also reflected in their performance. Selected exploration strategies are recorded in Figure~\ref{fig_exppin2}. 

\subsection{Rendering with Binary Layers: Performance}

For the accuracy and time-to-completion performance metrics, a Shapiro-Wilk test was used to evaluate the normality of the user scores for both display types, then a two-way repeated measures ANOVA was used to assess the impact of display and model type on participant performance when assembling each of the three models for each display type. A post-hoc test was used to evaluate the relative impact of each display type.

\begin{figure}[H]
\centering
\includegraphics{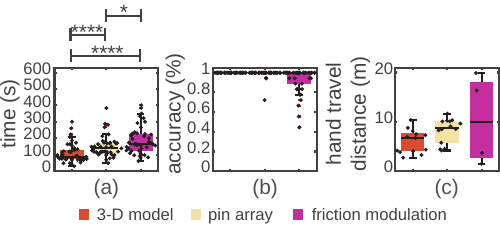}
\caption{\textbf{Performance Results for Binary Layers Assembly Task.} (a) The total time for a participant to complete each assembly task across all three configurations; (b) accuracy was measured by calculating the number of correctly placed blocks divided by the maximum possible score; and (c) total distance traveled by each of the hands for each display type. Data shown is only for participants where hand tracking was measured as 'successful'. * corresponds to p $<$ 0.05, **** corresponds to p $<$ 0.0001.}
\label{fig_BL All}
\end{figure}


\subsubsection{Accuracy for Binary Layer Assembly Task}
Accuracy was near the ceiling for most participants, so no ANOVA was conducted. The results are visualized in Figure~\ref{fig_BL All}(b).


\subsubsection{Time-to-Completion for Binary Layer Assembly Task} 

There was a significant difference in time between the 3-D model and the friction-modulating display (p=2.31E-06$<$0.05), the 3-D model and the pin array (p=8.82E-06$<$0.05), and between the pin array and the friction-modulating display (p=2.90E-02$<$0.05) for assembling the rendered models. Time-to-completion was at a minimum for the 3-D model, followed by the pin array. Time-to-completion performance was at its highest when using the friction-modulating display. 

Data were transformed with lambda = 0.05 prior to analysis.
The untransformed data is visualized in Figure~\ref{fig_BL All}(a).


\subsubsection{Energy Expenditure for Binary Layer Assembly Task.}

This was averaged across all three models. A one-way ANOVA was used to assess the impact of display configuration, which found that platform did not have a significant effect on energy expenditure (p=0.788$>$0.05). 
A Tukey Ladder of Powers produced a lambda = 0.275 for transforming data; 
untransformed data is visualized in Figure~\ref{fig_BL All}(c). 
Hand tracking again varied wildly between participants; as a result, trials where hand tracking was determined to have 'failed' (i.e., $<$20\% tracking success) were excluded from analysis. 

\subsection{Rendering with Binary Layers: User Experience}

\begin{figure}[H]
\centering
\includegraphics{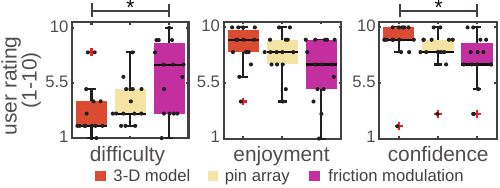}
\caption{\textbf{User Experience Ratings for Binary Layers Assembly Task.} Participants were asked to rate, on a scale of 1-10: (A) how difficult the assembly task was, (B) how much they enjoyed the assembly task, (C) how confident they were in interpreting and assembling the 3-D Model. * corresponds to p $<$ 0.05.}
\label{fig_BL Survey}
\end{figure}

A one-way ANOVA was used to assess the impact of display type on user experience, and a post-hoc paired sample t-test with a Bonferroni correction was used to assess the relative effect of each.
There was a significant difference in perceived task difficulty between the 3-D model and friction-modulating display (p=0.045$<$0.05), where the 3-D model was perceived as easier to use; but not between the 3-D model and pin array (p=0.139$>$0.05) or between the pin array and the friction-modulating display (p=0.092$>$0.05). There was not a significant difference in task enjoyment between the 3-D model and friction-modulating display (p=0.149$>$0.05), the 3-D model and pin array (p=0.984$>$0.05), or the pin array and friction-modulating display (p=0.192$>$0.05). There was a significant difference in confidence between the 3-D model and friction-modulating display (p=0.042$<$0.05), where confidence was lower for the friction-modulating display; but not between the 3-D model and pin array (p=0.242$>$0.05) or between the pin array and friction-modulating display (p=0.265$>$0.05).
A Tukey Ladder of Powers resulted in lambda = 0.175, 2.5, 3.5 for difficulty, enjoyment, and confidence. The untransformed data is visualized in Figure~\ref{fig_BL Survey}. 


\section{Discussion}
This work evaluated two tools for encoding 3-D surface information across four surface haptic display configurations and identified strategies to optimally leverage their information-encoding capabilities.
Collectively, our results highlight how surface haptic displays trade spatial resolution, interaction bandwidth, and exploratory freedom when encoding 3-D structure.

Overall, all four surface haptic display configurations were effective for communicating 3-D information using the height map method. While participants took significantly longer to interpret encodings on surface haptic displays compared to 3-D printed models, the combination of high accuracy and potential for performance improvement with exposure suggests that surface haptic displays provide a viable medium for 3-D data, free from the limitations inherent in non-digital tactile tools.

Modifications such as increasing pin density and dividing the friction-modulating display into distinct sections improved user experience and reduced completion time, though they had minimal effect on accuracy or energy expenditure. Participants preferred the high-density pin array to its low-density counterpart and found it easier to interpret the embedded reference models. Both friction-modulating screens were rated as more challenging than the 3-D printed model; however, the stenciled surface was perceived as comparably enjoyable, and participants reported greater confidence when using it compared to the non-stenciled screen. 
These findings suggest that relatively minor design adjustments can markedly enhance the effectiveness of haptic displays for conveying tactile information. Although direct comparisons between pin arrays and friction-modulating displays were not possible due to time constraints, their performance metrics were similarly promising. At the resolutions tested, the two technologies appear potentially interchangeable, with friction modulation becoming particularly advantageous when a larger number of distinct tactile keys is required.
These limitations motivate an alternative encoding strategy that reduces per-cell precision in favor of representing more complex 3-D information through sequential tactile layers.

Whereas height maps emphasize continuous depth cues at each spatial location, the binary-layer approach trades local resolution for increased representational flexibility.
The results for the binary layers interaction largely aligned with our expectations, with accuracy maintained at a high level, demonstrating, at a minimum, the feasibility of the assembly task. Notably, accuracy exceeded that of the height map task. For the pin array, this may be attributed to the ease with which participants could interpret adjacent matrix cells to collect height information; for the friction-modulating display, textures were more distinct with substantial differences in spatial frequency, reducing the cognitive load to distinguish between two textures. Participants required more time to interpret on-screen objects when using the friction-modulating interaction mode than with the alternative interaction modes, especially when compared to the 3-D printed models. Mean energy expenditure aligned with the time-to-completion metrics; however, due to limitations in available data, this could not be statistically verified. 

In terms of user experience, responses for the pin array closely matched those for the 3-D printed model across all survey metrics. The friction-modulating display had a wider variability of performances, was perceived as more difficult, and participants experienced lower confidence in interpreting the on-screen objects compared to the 3-D printed model. Sensation using friction modulation is subject to variation based on factors such as fingertip moisture~\cite{AliAbbasi2024}. These factors may have negatively impacted performance compared to the other two methods of encoding 3-D information, potentially adversely impacting performance in comparison to the other methods. Additionally, exploratory strategies on friction-modulating displays are more restrictive than those when exploring the 3-D model or pin array. The inclusion of a stencil, as employed in the "height map" experiments, or the integration of multi-touch capabilities, may reduce the time required for locating subsections and facilitate direct comparisons among multiple textures. 

All three tested displays are suitable for communicating 3-D information as a sequence of slices, particularly with increased user exposure. Friction modulation is especially promising due to its compatibility with flat mainstream touchscreen devices, offering a practical pathway for integrating tactile 3-D interaction into widely used interfaces.

\subsection{Study Limitations and Future Work} 
The effectiveness of the hand tracker varied considerably across participants, likely due to differences in hand size and display type. As a result, it was useful only for coarse comparisons between display configurations. Distinguishing the energy expended in interpreting the model from that expended in physically building it proved challenging; therefore, only aggregate hand movements were measured, and no statistical analysis was conducted.

Gaps between cells in the high-density pin array disrupted movement between the cells of the height map matrix, complicating some participants’ evaluation of relative height information. Participants also encountered difficulty locating and manipulating the magnetic blocks. An experimental setup with a more ergonomic design and clearer separation between assembly and exploration zones may better support sustained exploration and interpretation.

An additional limitation is that the pin array configurations were evaluated using static 3-D printed models rather than refreshable hardware. Although the low-density configuration was based on the dimensions and pin spacing of the commercial Graphiti display, the higher-density configuration was an experimental modification. Thus, the feasibility and performance of implementing these configurations on refreshable hardware remain to be established. Future work should evaluate these configurations on functional refreshable displays.

Although a training session was included, the short-term performance observed in these experiments does not predict long-term proficiency, given that this represents a novel method for most individuals for the purposes of interpreting 3-D information. Future studies might benefit from assessing participants’ prior experience with tactile tools and displays to contextualize their performance.

The findings of this study highlighted several promising directions for further research. One possible avenue involves exploring the effects of increased grid resolution on performance. Although other studies have showcased impressive 3-D modeling techniques and technologies using haptic feedback~\cite{Siu2019}, such displays still often fall short in effectively rendering internal geometries or features such as overhangs or cavities. Therefore, examining the extent to which pin arrays can be utilized for this purpose remains crucial. 

For friction-modulating displays, it is unclear whether higher resolutions can be achieved without making assembly tasks impractical. This uncertainty arises from two factors: (1) textures require a minimum level of exploration to identify them, so they may not be possible to scale down, and (2) in the absence of a stencil, users may struggle to gauge a stimulus's location in space. As a result, reconstructing higher-resolution 3-D graphics may become prohibitively difficult. Future research should aim to characterize these limitations and explore design solutions that improve display resolution or circumvent the need for higher display resolutions.

Finally, participants experienced difficulty discerning stimulus boundaries on the pin array. This highlights the need to develop methods for clearly demarcating or emphasizing shape features without disrupting the perceived continuity of a surface.

\section{Conclusion}
\begin{figure}[t]
\centering
\includegraphics{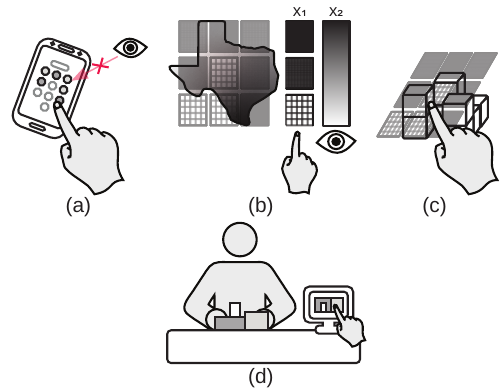}
\caption{\textbf{Potential Practical Applications of Conveying 3-D Information Using Tactile Feedback.} Example integrations for multi-modal visuo-haptic feedback on surface displays, including, but not limited to, (a) enhanced security integrations for password-protected devices, (b) increased information encoding potential for multi-modal graphics, (c) accessible 3-D graphics, and (d) accessible assembly instructions.}
\label{fig_appllications}
\end{figure}

The work presented two methods for conveying 3-D information via tactile feedback. We evaluated the effectiveness of these methods when applied to displays representative of existing refreshable haptic displays, with or without usability-oriented modifications. Both the height-map and binary-layer approaches proved effective across all display configurations, with users achieving high accuracy for both methods.

Minor modifications to each display type improved time efficiency and user experience for the height map method, highlighting opportunities for refining current-generation surface haptic display designs. One promising avenue for future exploration is the assessment of a multi-touch friction-modulating display, which could expand the range of possible exploratory strategies and overcome current limitations in interpreting on-screen graphics.

Performance differences between displays can be explained in part by the number of distinct “intensity” keys (e.g., pin heights or spatial frequencies). By limiting comparisons to two levels at a time, users more easily interpret the rendered information, explaining the pin array’s superior performance for the binary-layer method. Generally, both surface haptic displays underperformed relative to the 3-D prints; however, increased exposure and training may improve both accuracy and user experience.

These methods and displays are well-suited as multimodal design tools capable of operating independently of visual feedback, potentially enhancing engagement with 3-D graphics and improving accessibility. Practical applications include, but are not limited to, improved security tools, graphics with higher information encoding capacities, accessible 3-D graphics, and accessible assembly pipelines (see Figure~\ref{fig_appllications}). Moreover, integrating haptic feedback on surface displays addresses the high storage costs and inflexibility associated with traditional analog tactile tools. Future work should continue to explore optimal rendering strategies, higher-resolution displays, and techniques for supporting more complex 3-D structures.

\bibliography{references}
\bibliographystyle{unsrt}

\end{document}